%% file: main.tex
\documentclass[a4paper,fleqn,usenatbib]{mnras}

\include{definitions}

\title[Deep Learning model for LLAGNs emission]{Deep learning model for multiwavelength emission from low-luminosity active galactic nuclei}

\author[Almeida et al.]{
Ivan Almeida$^1$\thanks{E-mail: ivan.almeida003\@gmail.com}\orcid{https://orcid.org/0000-0001-6018-2852},
Roberta Duarte$^1$,
and Rodrigo Nemmen$^1\orcid{https://orcid.org/0000-0003-3956-0331}$ 
\\
$^1$Universidade de S\~ao Paulo, Instituto de Astronomia, Geof\'{\i}sica e Ci\^encias Atmosf\'ericas, Departamento de Astronomia,\\ S\~ao Paulo, SP 05508-090, Brazil\\
}

\date{Accepted XXX. Received YYY; in original form ZZZ}
\pubyear{2021}

\begin{document}
\label{firstpage}
\pagerange{\pageref{firstpage}--\pageref{lastpage}}
\maketitle

\begin{abstract}
Most active supermassive black holes (SMBH) in present-day galaxies are underfed and consist of low-luminosity active galactic nuclei (LLAGN). They have multiwavelength broadband spectral energy distributions (SED) dominated by non-thermal processes which are quite different from those of the brighter, more distant quasars. Modelling the observed SEDs of LLAGNs is currently challenging, given the large computational expenses required. In this work, we used machine learning (ML) methods to generate model SEDs and fit sparse observations of LLAGNs. Our ML model consisted of a neural network and reproduced with excellent precision the radio-to-X-rays emission from a radiatively inefficient accretion flow around a SMBH and a relativistic jet, at a small fraction of the computational cost. The ML method performs the fit $4 \times 10^5$ times faster than previous semianalytic models. As a proof-of-concept, we used the ML model to reproduce the SEDs of the LLAGNs M87, NGC 315 and NGC 4261.
\end{abstract}




\section{Introduction}\label{sec:introduction}

Quiescent galaxies with little or no ongoing star formation dominate the local universe \citep{Bell2004, Bundy2006, Faber2007, Ilbert2010}. The observed quiescence can be related to the activity phase of an active galactic nucleus (AGN). In neighbor galaxies with AGNs, we do not observe a relevant quasar population. Instead the observations show an active galaxy population mainly formed by low-luminosity AGNs (LLAGNs); some examples are SgrA* \citep{Narayan1995nat,Yuan2003}, NGC 3115 \citep{Wong2014}, and M87 \citep{EHTC2019}.

LLAGNs present different observational signatures compared to more luminous AGNs -- i.e., quasars, blazars. All AGNs display a broadband multiwavelength emission, but LLAGNs did not have prominent UV continuum emission -- characteristic of an optically thick and geometrically thin accretion disc \citep{Ho1999, Ho2008, Nemmen2006, Wu2007}. LLAGNs show weak and narrow Fe K$\alpha$ \citep{Terashima2002} lines in their spectra, which agrees with a thin accretion disc absence. Furthermore, the observed luminosity from these LLAGNs is far below the expectation for a thin disc with an efficiency of 10\% ($L = 0.1\dot{M}c^2$; being $L$ the luminosity and $\dot{M}$ the accretion rate). These characteristics suggest that LLAGNs are radiatively inefficient systems drastically different from the standard scenario.

The AGN dynamics depends on how the supermassive black hole (SMBH) accrete the available gas in its surroundings. These processes rely strongly on whether the viscously generated thermal energy is radiated away \citep{Abramowicz2013}. The LLAGNs in our nearby vicinity are underfed SMBHs with accretion rate $\dot{M} < 10^{-2} \dot{M}_{\rm Edd}$ ($\dot{M}_{\rm Edd}$ is the Eddington accretion rate), these objects are classified as radiatively inefficient accretion flows (RIAF). RIAFs are extremely hot, optically thin, and geometrically thick accretion discs; for more details, see \cite{Yuan2014}.

RIAFs have characteristic emission in two primary wavelengths: radio and X-ray. The radio component of the spectral energy density (SED) is related to synchrotron emission in the accretion disc. The elevated temperatures in RIAFs \citep{Yuan2014} indicate an electronic population with very high average energy. These electrons can interact with available photons and transfer energy via inverse Compton scattering (IC). IC is one of the primary sources of X-ray emission. RIAFs can emit bremsstrahlung radiation also, via electron-ion scattering inside the accretion flow. However the radiation field in RIAFs is very faint, and it has negligible effects in the optically thin accretion flow dynamics.

An astrophysical object's emitted light carries information about the physical processes occurring in the source. With the observed SED, one can model the emitting source. LLAGNs have a broadband emission, and modeling their multiwavelength emission gives us valuable knowledge about the system -- e.g., the accretion rate or electronic and ionic distributions in the accretion flow. Several works \citep{Yuan2003, Nemmen2006, Nemmen2014, Almeida2018, Bandyopadhyay2019} performed similar analyses. Traditional methods to fit LLAGN SEDs (e.g. \cite{Nemmen2014}) are computationally intensive and require the user to actively monitor and participate in the modelling process.

In the last few years, machine learning (ML) methods have become extremely common in science, especially in dealing with big data problems, like Astronomy. The development of better telescopes gives us a massive amount of data, which is a favorable scenario to implement ML algorithms \citep{George}. Many authors published works using ML methods in astronomical data, doing object classification \citep{Rhode2005, Banerji2010}, and spectra analysis and modeling \citep{Firth2003, Valdes2006, Ball2007, marul2020, liewcain2020}, for example. 

Neural Network (NN) is a ML method based on biological neurons and how they pass through the information \citep{McCulloch1943}. It is widely used in the scientific community since they show results when it comes to classification \citep{Wan1990, storrie1992, odewahn1995} and regression \citep{specht1991, comrie1997}. NNs consist of neurons and connections between them. With weights ($w_i$) associated with each connection. Each neuron has a activation function $f(\sum(w_ix_i))$ that is a relation between the inputs ($x_i$) and weights \citep{leshno1993}. Deep learning (DL) is a very powerful tool, for more details see \citep{Breen2020, Rodriguez2018, Lecun2015}.

Other works used NNs to fit observational data  \citep{Asensio2009, Pacheco2019, fathivavsari2020}. For instance,  \cite{Asensio2009} shows a technique based on the combination of two ML methods: principal analysis component (PCA) and NNs.  It uses a PCA for dimensionality reduction, and the NNs are used to interpolation. The PCA is useful to get faster results since it can shrink the size of the data. In our work, we do not perform a dimensionality reduction, but the speed-up is adequate. \cite{fathivavsari2020} proposed the use of NNs as a new technique to predict the flux and the shape of Ly$\alpha$ emission lines in the spectra from quasars. The architecture proposed in \cite{fathivavsari2020} is a deep NNs with five layers with the input being $Si_{IV}$, $C_{IV}$, and  $C_{III]}$ emission lines, and the output is the Ly$\alpha$ emission line. Similarly to our work, they used hyperparameter tuning methods to find the best architecture. They obtained a NN able to predict  Ly$\alpha$ emission lines with a precision $6\%-12\%$.

We aimed at calculate LLAGNs' SEDs using deep learning methods. To model the observational data, we needed to generate a large number of SEDs to find the best fit via traditional methods --like the minimum square error. Creating a large amount of SEDs is challenging; the computational time involved is very high. We used DL method to speed-up our SED generation rate up to 100000 times, depending on the spectrum process. For this work, we generated a training sample of $\sim20000$ SEDs for RIAFs and $\sim 140000$ for jets changing the main parameters of the system, and we trained a neural network (NN) with the dataset. The code's name is \texttt{AGNNES} (\textbf{A}ctive \textbf{G}alactic \textbf{N}uclei \textbf{Ne}ural network \textbf{S}ED generator).

In section \ref{sec:model}, we presented the details of the model, details about the NN construction and training are in section \ref{sec:neural-network}. In section \ref{sec:results} we presented our results. We compared our results to other similar works in literature in section \ref{sec:discussion}. The summary of the work is in section \ref{sec:summary}.

\section{Modelling}\label{sec:model}

\subsection{Original data}\label{subsec:data}

We assumed sub-Eddington LLAGNs are accreting as radiatively inefficient accretion flow (RIAF). We used a semi-analytical approach to treat the RIAF emitted radiation \citep{Nemmen2014}. Normally LLAGN SEDs are radio-loud \citep{Ho1999, Sikora2007}. In our model, we also included the synchrotron emission from a relativistic jet, either. Here in this work, we followed the same approach as \cite{Nemmen2014} to generate our SED sample. \footnote{Fiducial model source code can be found in: \url{https://bitbucket.org/nemmen/adaf-code/src/master/}}.
Henceforth we will call this model as the ``fiducial'' one.

\subsubsection{RIAF}\label{subsubsec:RIAF}

We considered the calculation of an optically thin and geometrically thick two-temperature accretion flow with outer radius $r_0 = 10^4R_S$ ($R_S$ is the Schwarzschild radius), which presents very low radiative efficiency \citep{Narayan1998}. Our assumptions to calculate the system SED were:
\begin{enumerate}
    \item The accretion flow is stationary.
    \item Viscosity is parameterized as \cite{Shakura1973} $\alpha$-viscosity.
    \item The gravity is described with a pseudo-Newtonian potential \citep{Paczynsky1980}.
\end{enumerate}
The radiative transfer is treated carefully in more detail. We took into account radiation emission from synchrotron, inverse Compton scattering (IC), and bremsstrahlung processes occurring inside the accretion flow. \citep{Nemmen2006, Yu2011, Nemmen2014}. 

Outflows, or winds, are an intrinsic feature of RIAF's model, e.g. \cite{Yuan2012, Yuan2015, Almeida2020}. In this work, we considered the parameterization proposed by \cite{Blandford1999} for the accretion rate as a function of radius following a power-law relation:

\begin{equation}
    \dot{M} = \dot{M}_0 \left( \frac{r}{r_0} \right)^s
\label{s-equation}
\end{equation}
the parameter $s$ is related to the ``intensity'' of winds, higher values of $s$ mean stronger winds. $\dot{M}_0$ is the accretion rate at the defined radius value $r_0$ in units of $\dot{M}_{\rm Edd}$. In this work, we used $r_0$ as the outer radius of the accretion disc, $r_0=10^4 R_S$. Following \cite{Nemmen2014} we assumed that $s$ is limited to the range $0 \lesssim s \lesssim 1$. For $\dot{M}_0$, we considered a sub-Eddington system and limited the possible values to $\dot{M}_0 < 0.1\dot{M}_{\rm Edd}$.

The RIAF solution depends on the system's physical parameters: the black hole mass $M$, the viscosity parameter $\alpha$, the adiabatic index $\gamma$, the parameter $\beta$ -- called modified plasma parameter -- defined as the ratio between gas pressure and total pressure ($\beta = P_{\rm gas}/P_{\rm total}$), and the fraction of turbulence energy dissipated that heats the electronic population of the plasma $\delta$. Here we followed \cite{Nemmen2006} and adopted $\alpha = 0.3$, $\beta =0.9$ and $\gamma = 1.5$. For $\delta$ values, normally they were considered small ($0.01$ in \cite{Narayan1995}), but others processes could increase this value \citep{Quataert1999turb, Sharma2007}, like magnetic reconnection in hot plasmas for example, which can allow $\delta$ to vary between $0.01 \leq \delta \leq 0.5$. 

In the RIAF model, we varied three free parameters: $\delta$, $s$ and $\dot{M}_0$. We generated $\sim 20000$ RIAF SEDs randomly sampling the parameters in range: $0.01 \leq \delta \leq 0.3$; $0 \leq s \leq 1$; $-5 \leq \log_{10} \left(\dot{M}_0/\dot{M}_{\rm Edd} \right) \leq -1$.

\subsubsection{Jet}\label{subsubsec:jet}
The RIAF component does not produce enough radio emission to match the radio observations of LLAGN \citep{Ulvestad2001, Nemmen2006, Liu2013}. We included in our model the contribution of a jet component modeled based on the scenario of internal shocks \citep{Spada2001}. Following this model in the SMBH surroundings, a portion of the accretion flow gas is transferred to the jet generating a mass outflow rate $\dot{M}_{j}$ and a shock wave. The shock wave makes the jet emission dominated by a non-thermal leptonic population.

Our modeled jet has a conical geometry with a half-opening angle of 0.1 rad ($\sim 6^{\circ}$) and a constant bulk Lorentz factor of 2.9, independent of the distance to the central SMBH. The jet is perpendicular to the accretion disc with an angle with the line of sight of $30^{\circ}$. The shocks accelerate the electrons to a power-law with index $p$. The parameters $\epsilon_e$ and $\epsilon_B$ describe respectively the fraction of energy density from the electrons and the magnetic field.

The jet modelling free parameters were: $\dot{M}_j$, $p$, $\epsilon_e$ and $\epsilon_B$. We allow $p$ to be in the range $[2,3]$, as the shock theory predicted it. We expect that $\dot{M}_j < \dot{M}_0$, here $\dot{M}_0$ is from RIAF modeling, the mass outflow in the jet should not be higher than the mass inflow in the accretion disc. By definition, both $\epsilon < 1$, since they are fractions of the total energy. We generated $\sim 140000$ jet SEDs. We randomly varied the free parameters in the range: $2 \leq p \leq 3$; $-7 \leq \log_{10} \left(\dot{M}_j/\dot{M}_{\rm Edd} \right) \leq -2$; $-5 \leq \log_{10}\epsilon_e \leq -1$; $-5 \leq \log_{10}\epsilon_B \leq -1$.

\subsection{Neural network}\label{sec:neural-network}
Our model is a deep neural network composed of neurons with several layers and weights. In the output layer, we have a function called loss function $\mathcal{L}$ which is any error function that gives us how much the prediction and target are different. The learning procedure is given by the backpropagation method that consists of derivatives of $\mathcal{L}$ concerning the weight we want to learn \citep{kelley1960}. In other words, the backpropagation uses gradient descent \citep{ruder2016} to find the minimum value in a space given by $\mathcal{L}(w_{ij})$, i.e., find which weights $w_{ij}$ gives the minimum of $\mathcal{L}$. 

The learning part starts when we feed the neural network with data from the training set. We can give the data in batches, which is giving data by parts; this corresponds with faster training. Epoch is passing all batches once. The batch size and epoch are called hyperparameters. They are changed to find which combination of hyperparameters gives better results. In our work, we considered the number of neurons and the number of layers as hyperparameters. In this way, our goal was to find the best architecture for our model. We created a \textit{GridSearch} to find the best values of our hyperparameters. 
The \textit{GridSearch} is a hyperparameter tuning method \citep{Bergstra2012} that consists of train the NN with several combinations of hyperparameters. In our case, our hyperparameter is the number of neurons in each layer. We created a shell script that runs all the combinations possible and saves the chosen metric. The trainings with the better metrics are the one we use to further analyze. Our best model is composed of two architectures, one is to analyze the RIAF and the other to analyze the jet region. The RIAF model is composed of 4 layers with 56, 60, 99, and 99 neurons, respectively. The jet model is composed of 5 layers with 10, 44, 66, 99, and 130 neurons. Each layer has a ReLU activation except for the last layer which is a linear function. The mean absolute error as $\mathcal{L}$ and the optimizer is Adam \citep{Kingma2014}. 

To summarize, we build our neural network from scratch using the \textit{GridSearch} technique. Our architecture is composed of hidden layers meaning that it is a deep neural network. We tested other architectures with different hyperparameters, but this is the best model. 

\subsection{AGNNES's Performance}\label{subsec:performance}

SED calculations for LLAGN are time expensive. The fiducial code which we generated our training RIAF SED's sample took a minute per spectrum\footnote{Depends on the hardware and parameters set. For our available computers, it took 0.5-2min}. The trained NN, AGNNES, calculates the same component in approximately $0.3$ ms. It is $\sim 400000$ times faster than the original RIAF code. The time improvement allowed us to implement more robust statistic methods to LLAGN SED fitting, which is absent in previous works. The original jet calculation is much faster than the RIAF component, took around $\sim 0.2$ seconds each jet SED. AGNNES calculates the same component in approximately $0.4$ ms. The improvement, for the jet, is close to $\sim 500$ times. 

\begin{figure}
    \center
    \subfigure[RIAF]{\includegraphics[width=\linewidth]{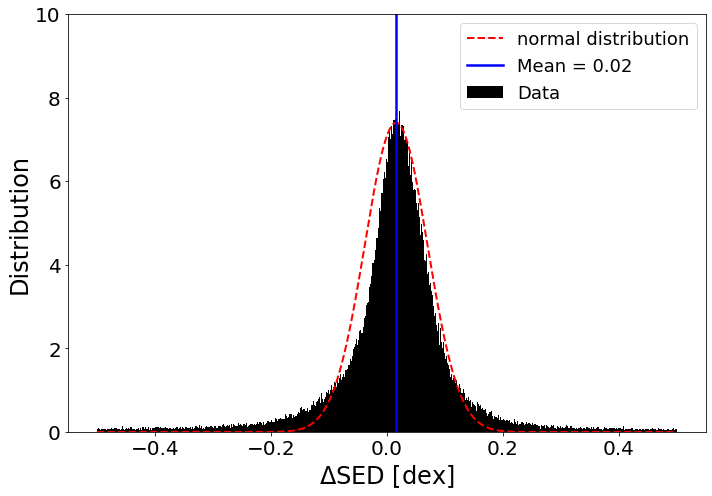}}
    \qquad
    \subfigure[Jet]{\includegraphics[width=\linewidth]{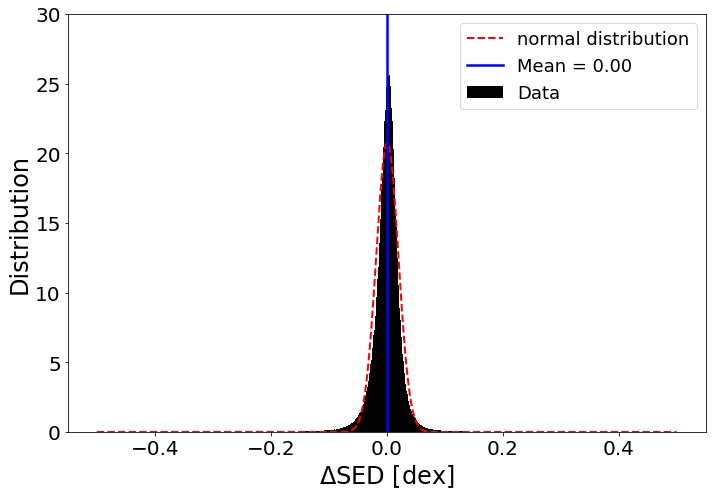}}
    \caption{From the validation set, we calculated the difference between $\log_{10} \nu L_{\nu}$ from the original data and the calculated by AGNNES and plotted these values as a histogram. We obtained $\overline{\Delta {\rm SED}} \approx 0.02 \pm 0.05$ for RIAF model and $\overline{\Delta {\rm SED}} \approx 0.00 \pm 0.01$ for jet model, respectively top and bottom panels.}
    \label{fig:nn-error-RIAF}
\end{figure}

AGNNES results are an approximation from the original calculation. We had an extremely high speed-up in calculations, but the NN introduces inherently small errors when compared to the fiducial code. AGNNES, on average, can reproduce very well the validation data -- see appendix \ref{app:nn}. Comparing the AGNNES predictions with the validation data sample, we defined $\Delta {\rm SED}$ as the ``distance'' between the model and AGNNES prediction:

\begin{equation}
     \Delta {\rm SED}(\nu) \equiv \log_{10} ( \nu L_{\nu})_{\rm original} -  \log_{10}( \nu L_{\nu})_{\rm AGNNES}
     \label{eq:delta-sed}
\end{equation}
 Considering the validation sample and comparing original data against AGNNES predicted one, for every point (i.e. frequency). We found the averaged value of $\overline{\Delta {\rm SED}}$ = $0.02 \pm 0.05$ (see \eqref{eq:delta-sed}), we plotted in figure \ref{fig:nn-error-RIAF}. For the RIAF component the NN predicted slightly smaller emission values. The jet component presented $\overline{\Delta {\rm SED}}$ = $0.00 \pm 0.01$. These $\overline{\Delta {\rm SED}}$ values are AGNNES's uncertainty.

The original data is not perfect. Sometimes the original code randomly fails while calculates the inverse Compton emission, returning a lower value for the emission. We automatically generated the data, and we identified the code failures and discarded the wrong data automatically, but it was not 100\% efficient. The data set have few``bad'' SEDs mixed in the whole set; this can affect the NN training and be responsible for the slightly systematic deviation of 0.02 dex in $log_{10}(\nu L_{\nu})$. However, LLAGNs are very difficult to observe, and the observations typically have considerable uncertainties, higher than 0.02 dex. The theoretical models for the LLAGN emission have their uncertainties. The NN averaged deviation from the original model is negligible if compared with the usual error bars in measurements and model uncertainties.

\subsection{Fitting Method}\label{subsec:fitting}

The current approach for LLAGN SED fitting in literature is an iterative method in which one changes all the parameters individually while the others remain unchanged. The best fit is determined visually after a certain number of attempts with no estimated uncertainties on the results \citep{Nemmen2014, Almeida2018}. The SED calculation is computationally expensive. The bottleneck of LLAGN SED calculation is in the Comptonization. The fiducial code takes $\sim 2$min to generate a single SED. It is tough to explore a vast range of parameters; there is insufficient time for this. For example, to run $10^5$ SEDs, we would need approximately two months without pause. With a faster method to calculate SEDs, we could explore a broader range of parameters and apply more robust fitting methods.  

Time performance is the strongest point of AGNNES when we compare it with the iterative method. A well trained NN can do the IC step smoothly without high time spending. The NN is not a perfect copy of the original model. During the training, there is error propagation, and the NN final result is an approximation of the real value; for simplicity, we assumed the original code gives the perfect SED value, and we compared AGNNES with it. We discussed the NN performance in section \ref{subsec:performance}.

We implemented the fitting procedure using Markov-chain Monte Carlo method. We performed the chain calculation with the \texttt{Python} package \texttt{emcee} \citep{Foreman-Mackey2013}. We used \texttt{emcee} to estimate the posterior distributions for the parameters -- $\delta$, $s$, $\dot{M}_0$, $p$, $\dot{M}_j$, $\epsilon_e$, $\epsilon_B$ -- that better describe the data. We defined the likelihood function as a pure Gaussian likelihood. The final likelihood is the sum of all seven parameters' likelihoods. The priors of parameters have initially been the limits of the data set presented in sections \ref{subsubsec:RIAF} and \ref{subsubsec:jet}.

We started the MCMC with a flat distribution ball around random initial values with small radius, considering all parameters as adimensional. Our MCMC chain ran with 300 walkers for the number of steps N: $N>30000$. We neglected the first 20\% steps as burn-in. The value of N depends strongly on the used dataset. We estimated the MCMC convergence using the integrated autocorrelation time ($\tau$) implemented in \texttt{emcee} following \cite{Goodman2010} method; it should be $N/50 \gtrsim \tau$.

\section{Results} \label{sec:results}

We chose some objects with available SED data in the literature to fit. We aimed to find the best constraints for the accretion flow parameters that reproduce sources' SED. Our chosen objects were:  M87, NGC 4261, and NGC 315. The data points of all SEDs are available in appendix \ref{app:obs-data}.

To obtain the fit, we used existing independent measurements and theoretical models as \textit{priors}. In our work, we considered all our priors as flat distributions, for simplicity. For the SEDs plots, we adopted the following convention: The red dashed line is the RIAF contribution, the blue dash-dotted line is the jet contribution, and the solid black line is the sum of them. The grey shaded area is a set of one hundred curves generated by the MCMC method and represents the uncertainties in the total sum. We showed the posterior distribution of the free parameters in appendix \ref{app:distro}. 

We showed the fitting results in table \ref{tab:results}. The columns show the seven free parameters for RIAF and jet models, and the last one is the reduced $\chi^2$ calculated between the best fit and the observational data points. For observational points without uncertainty, we assume an uncertainty of 0.05 dex. 

\begin{table*}
    \centering
    \begin{tabular}{lcccccccc}
        \hline
        Object & $\delta$ & $\dot{M}_0$ ($\dot{M}_{\rm Edd}$) & $s$ & $\dot{M}_j$ ($\dot{M}_{\rm Edd}$) & $p$ & $\epsilon_e$ & $\epsilon_e$ & $\chi^2_{red}$\\
        \hline
        \\
        M87   &   --  &   --    &   --   &   $6.6^{+0.7}_{-0.8} \times 10^{-7}$    &   $2.54^{+0.04}_{-0.04}$   &   $4.7^{+0.4}_{-0.3} \times 10^{-2}$    &   $1.2^{+0.3}_{-0.2} \times 10^{-5}$  & $10.0$ \\
        \\
        NGC 4261    &   $0.30^{+2 \times 10^{-4}}_{-3 \times 10^{-4}}$  &   $9.97^{+0.02}_{-0.05} \times 10^{-3}$    &   $0.613^{+0.003}_{-0.003}$   &   $3.5^{+7.0}_{-7.5} \times 10^{-5}$    &   $2.4^{+0.4}_{-0.3}$   &   $2.4^{+3.8}_{-3.4} \times 10^{-4}$    &   $1.7^{+4.3}_{-6.6} \times 10^{-4}$  & $57.8$ \\
        \\
        NGC 315    &   $0.30^{+1 \times 10^{-4}}_{-3 \times 10^{-4}}$  &   $9.98^{+0.01}_{-0.03} \times 10^{-3}$    &   $0.507^{+0.003}_{-0.003}$   &   $4.7^{+0.9}_{-1.0} \times 10^{-5}$    &   $2.6^{+0.2}_{-0.1}$   &   $3.7^{+1.1}_{-1.1} \times 10^{-3}$    &   $1.2^{+0.3}_{-0.2} \times 10^{-5}$  & $6.7$ \\
        \hline
    \end{tabular}
    \caption{The final results of AGNNES's fit to our galaxy sample.}
    \label{tab:results}
\end{table*}

In this work, we favored the scenario in which the RIAF dominates the high energy emission. We did not consider IC emission from jet photons. The jet contribution was considered relevant mainly for low frequencies-- up to near-infrared --but neglected in higher frequencies for the galaxies NGC 4261 and NGC 315. In our jet fitting, we considered data points above infrared as upper limits for the jet emission. For M87 we modeled the whole emission favoring the jet model due to known observational priors \citep{Doeleman2012, Kuo2014}.

\subsection{M87}

\begin{figure}
    \centering
    \includegraphics[trim={0cm 0.cm 0cm 1.1cm},clip,width=\linewidth]{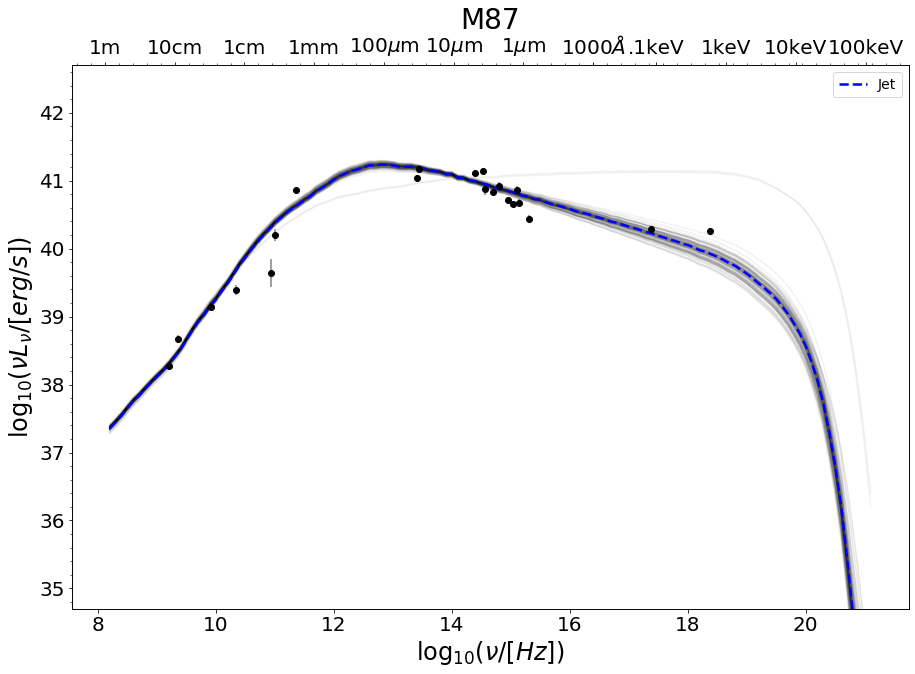}
    \caption{M87 SED best fit.}    \label{fig:M87-sed}
\end{figure}

The elliptical galaxy M87 harbors a supermassive black hole of $6.5 \times 10^9 M_{\odot}$ at 16 Mpc of distance from us \citep{EHTC2019}. The source presents strong radio emission due to a prominent relativistic jet. We showed the SED data points and the best fit in figure \ref{fig:M87-sed}.

For this source, we used as data points the high-resolution observations presented in \cite{Prieto2016} (c.f. Table 4). We treated observations with the lower spatial resolution-- presented in the same paper for different wavelengths --as upper limits.

The jet emission is a known feature of M87 \citep{Doeleman2012}. We tested models considering RIAF and Jet emission, we found the best result considering only the jet emission. Our fit suggested jet dominance over the inner RIAF for all wavelengths. It was not possible to produce a good fit for the observational SED using the RIAF model for this source. Our result considered only jet synchrotron emission and did not reproduce the bump around 100GHz and the most energetic data point is consistent with the best fit with $3\sigma$.

A thermal population of electrons in the accretion flow cannot reproduce the SED of M87 with such accretion rate. It is necessary to change the electronic distribution or assume the jet dominates the observed energy output entirely. The results for M87 fitting parameters are in table \ref{tab:results}, and we only considered the synchrotron jet in our model.

\subsection{NGC 4261 and NGC 315}

\begin{figure}
    \centering
    \includegraphics[trim={0cm 0.cm 0cm 1.1cm},clip,width=\linewidth]{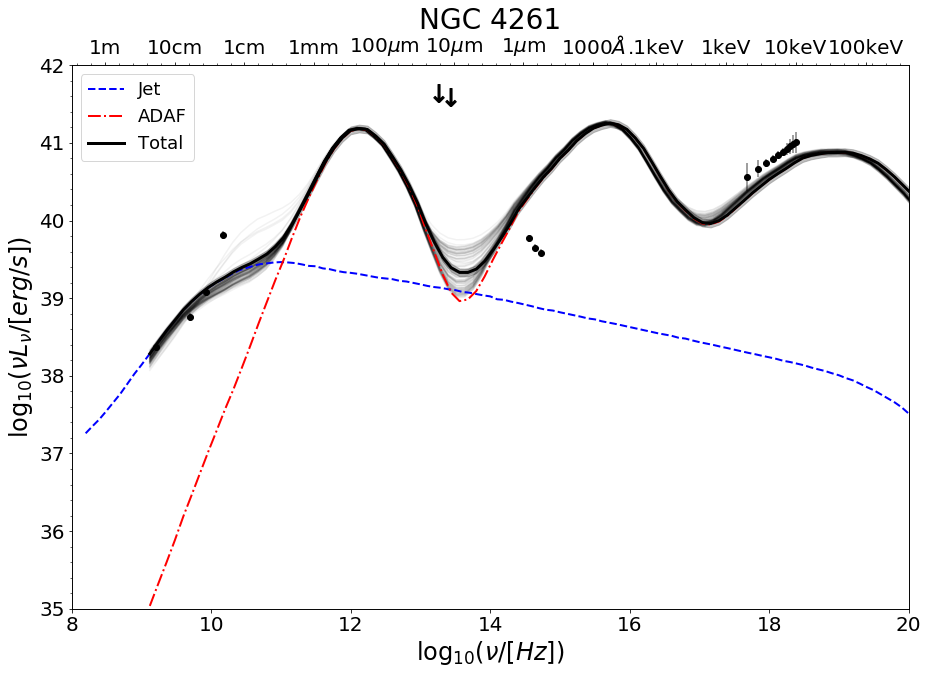}
    \caption{NGC 4261 SED best fit.}
    \label{fig:4261-sed}
\end{figure}

NGC 4261 is an elliptic galaxy at $\sim 30$Mpc of distance, located at the Virgo cluster. It harbors a SMBH with $5.2 \times 10^8 M_{\odot}$ \citep{Tremaine2002}. The SED data is available in table \ref{tab:NGC4261}. \cite{DeMenezes2020} presented data points and similar modeling for the same galaxy. We showed the SED data points and our best fit in figure \ref{fig:4261-sed}.

\begin{figure}
    \centering
    \includegraphics[trim={0cm 0cm 0cm 1.1cm},clip,width=\linewidth]{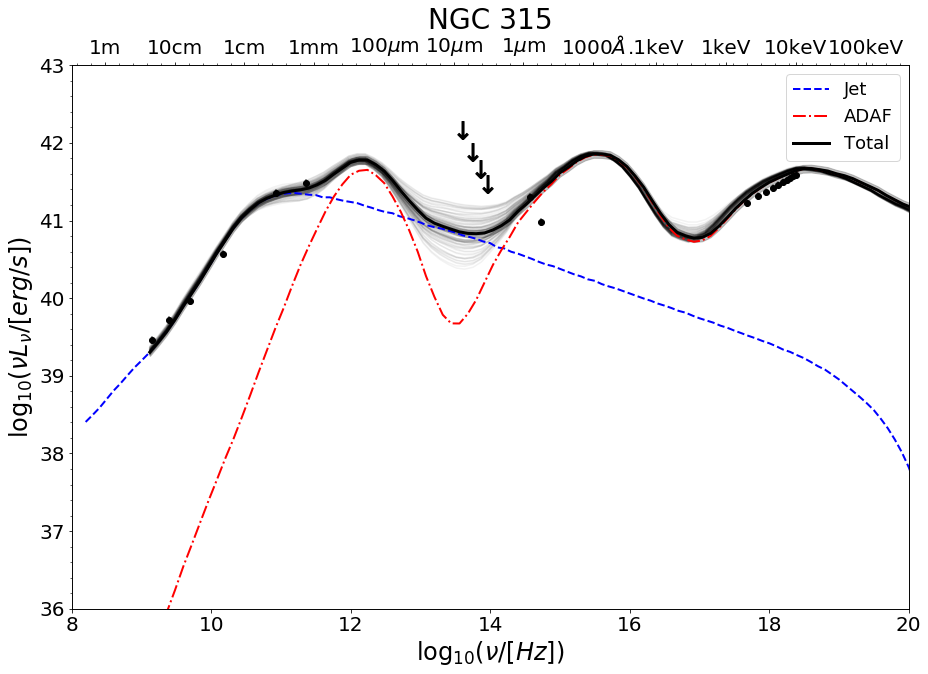}
    \caption{NGC 315 SED best fit.}
    \label{fig:315-sed}
\end{figure}

NGC 315 is another elliptical galaxy at $\sim 64$Mpc of distance, located at Pisces constellation. It contains a SMBH of $7.8 \times 10^8 M_{\odot}$ \citep{Woo2002}. The SED data is available in table \ref{tab:NGC315}. \cite{DeMenezes2020} presented data points and similar modeling for the same galaxy. NGC 315's SED data points and the best fit are in figure \ref{fig:315-sed}.

For these sources, we considered the optical data as upper limits. In the optical frequency range, there is a high possibility of stellar contamination. Stellar clusters typically inhabit the center of galaxies, and these populations emit a considerable amount of light in the optical band \citep{Lauer2005}. Since our primary interest here is the AGN emission, we neglected the potentially contaminated points by stars.

For NGC 4261 and NGC 315, our models were more focused on the RIAF component, especially for higher frequencies. In our model, the innermost regions of the RIAF produce most of the X-ray emission. We used the jet component to explain mostly the radio region of the SED. We did not consider the jet emission to explain the observed high energy emission (optical, UV, or X-ray bands). If we try to explain the whole SED only with a synchrotron jet, we will get different results and higher values of $\chi^2_{\rm red}$. 

For both NGC 4261 and NGC 315,-- assuming the RIAF dominates high energy emission --there are no strong constraints in the high-frequency jet emission ($\nu > 10^{12}\text{Hz}$) --, this assumption makes the value of $p$ poorly constrained. The jet model fits the radio points, and this gives us information about the accretion rate $\dot{M}_j$ and the distribution of energy, represented by $\epsilon_e$ and $\epsilon_B$. In our model, we restrained $\dot{M}_0 < 10^{-2} \dot{M}_{\rm Edd}$. We trained AGNNES to calculate SEDs with accretion rates up to $10^{-1} \dot{M}_{\rm Edd}$, but we expect even lower accretion rates for LLAGNs. Our assumption of the maximum value for $\dot{M}_0$ resulted in the truncated distributions for $\dot{M}_0$ in NGC 315 and NGC 4261, respectively figures \ref{fig:315-fit} and \ref{fig:4261-fit}. 

If the sources have higher accretion rates, the RIAF assumption is not valid. This truncation can be a hint that the emission could not come from an inner hot accretion disc. The X-ray emission for NGC 4261 could be better constrained if we allow an accretion rate of $\dot{M}_0 = 3.2 \times 10^{-2} \dot{M}_{\rm Edd}$.

For NGC 4261 and NGC 315, the fitting was not able to fully explain the UV emission. Both galaxies presented much lower observational value in UV than our RIAF model prediction.

\section{Discussion}\label{sec:discussion}

\begin{figure*}
    \center
    \subfigure[][NGC 4261]{\includegraphics[width=.47\linewidth]{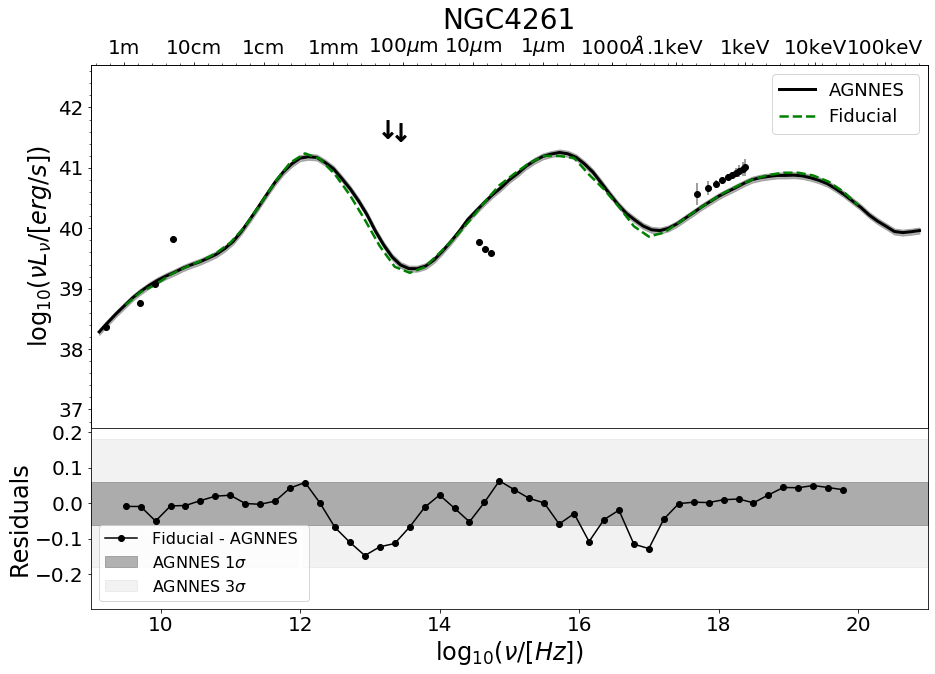}}
    \subfigure[][NGC 315]{\includegraphics[width=.47\linewidth]{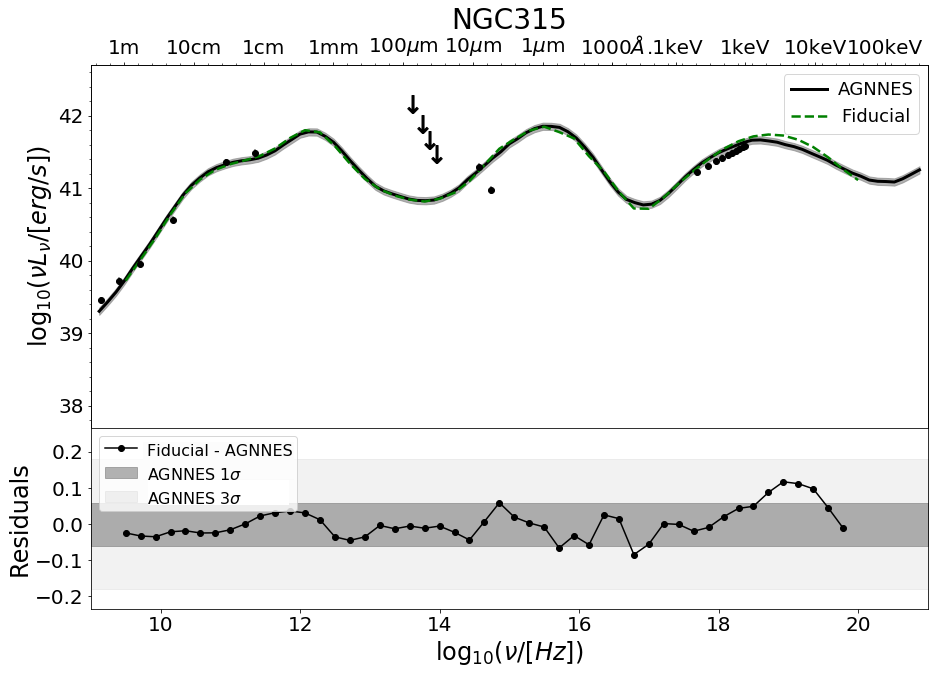}}
    \qquad
    \subfigure[][M87]{\includegraphics[width=.47\linewidth]{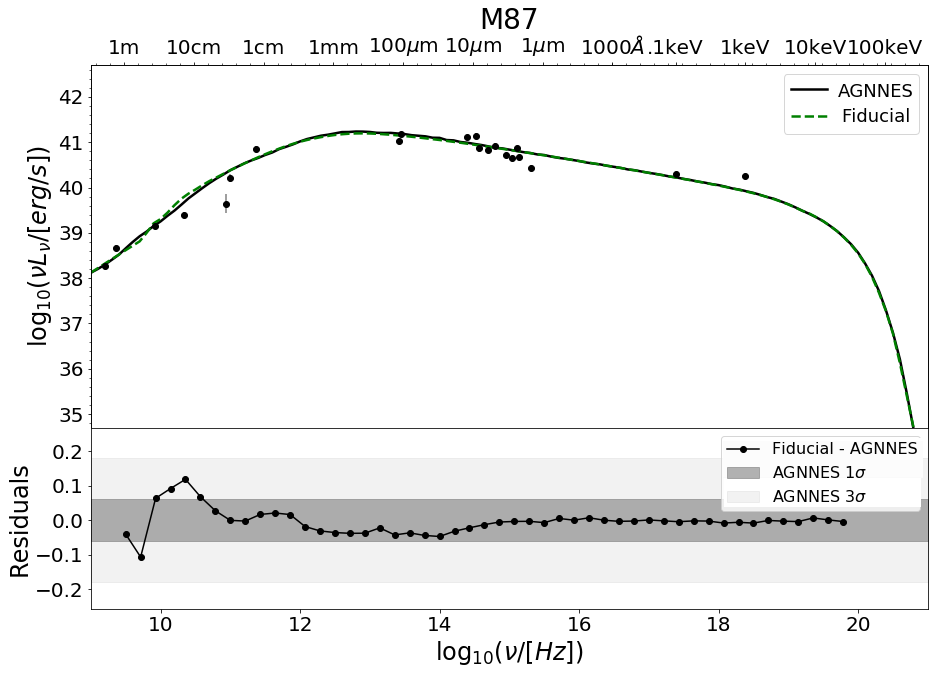}}
    \caption{Top panel: Comparison between SED calculated with AGNNES (solid black line) and following the original code (dashed green line). Bottom panel: The residuals between the original code SED and AGNNES's prediction. The darker grey region is $1\sigma$ uncertainty on the AGNNES result; the lighter grey region is $3\sigma$ region.}
    \label{fig:comparison-sed}
\end{figure*}

AGNNES's accuracy is an essential feature of this work. Using the parameters set calculated by AGNNES as the best fit, we plotted figure \ref{fig:comparison-sed}. In the top panel of this figure, the solid black line is AGNNES SED fit, and the dashed green line is the SED calculated with the fiducial code -- for the same parameters. For all sources, both curves are very similar. In the bottom panel, there are the residuals between the two curves. The grey zones represent the error of AGNNES for the uncertainty of $1\sigma$ and $3\sigma$ (see figure \ref{fig:nn-error-RIAF}). Considering all frequencies, from radio to X-ray, the fiducial code and AGNNES are equivalent. 

\subsection{Comparison with previous models}\label{subsec:observations}

Astronomers have studied M87 extensively over the past sixty years \citep{Felten1968, Schreier1982, Biretta1999, Doeleman2012, EHTC2019}. Recently, \cite{Bandyopadhyay2019} modeled this source with a combination of RIAF and synchrotron jet. They used a population of nonthermal electrons, which is absent in our work, and they manage to fit the bump at $\sim 100$GHz combining RIAF and jet emission. If they considered only thermal electron RIAF, the disc emission is not strong enough compared to the jet emission. 

For NGC 4261 and NGC 315, \cite{DeMenezes2020} did the previous SED modeling for the same data set with our original code with iterative procedures. In this work, we reported an improved version of these fits with AGNNES. Our results suggested that $\dot{M}_0 \gtrsim 10^{-2} \dot{M}_{\rm Edd}$, which is not consistent with the RIAF scenario -- notwithstanding our model fits well the data. \cite{DeMenezes2020} Gamma-ray observations indicate the emission comes from a jet emitting synchrotron (low-frequency) and synchrotron self-Compton (high-frequency).

\subsection{Model limitations}\label{subsec:model-limits}

The original model used to calculate the SEDs has its limitations, and AGNNES inherited all of them. The RIAF SED followed a straightforward model, with some strong approximations. First of all, it is a 1-dimensional calculation. It was assumed an axisymmetric accretion flow, and we integrated over the $z$-component (considering a cylindrical geometry). Furthermore, the code works on stationary accretion flow, hence it cannot account time variability.

Our jet SED took account of synchrotron emission only. We did not calculate the Inverse Compton effect in the jet photons. Some observations support a scenario with synchrotron self-Compton emission in the jet dominating the LLAGN emission \citep{Nagar2005, Finke2008, Takami2011}. 

The SED data of every object was not simultaneously measured. Many authors observed each data point at a different time, with years of separation. Moreover, we did not take into account any variability. We assumed all data comes from the quiescent state of the source. Such assumptions can be a problem for galaxies with few data points and not known variability patterns. M87 has more reliable data because it is a source continuously observed.

AGNNES works properly inside its training parameters range (see section \ref{subsec:data}). For fainter sources with very low luminosity, like SgrA*, AGNNES can not fit the SED reliably. For the SgrA* case, the accretion is too small to be inside our training specifications. The fiducial code can calculate SEDs for objects with such low accretion rates. However, numerical errors increase considerably as the accretion rate decreases. To generate data for low accretion rates are harder than for mildly higher accretion rates. The difficulty in creating SEDs with $\dot{M}_0 \gtrsim 10^{-5}$ impacted on AGNNES results. Below this $\dot{M}_0$ value, our NN can not calculate a correct SED. AGNNES best reproduces LLAGN SEDs with $(\nu L_{\nu})_{\rm peak} > 10^{36}$ erg/s.

\section{Summary} \label{sec:summary}

The primary objective of this work was to optimize the calculations of LLAGN SEDs, and the modelling of observed broadband spectra. Several previoues works used the iterative method to calculate the radiative emission from RIAF and jet for LLAGNs. The fiducial code spent few minutes to generate \textit{one} SED with two components.  To implement a more robust statistical analysis, we need a faster approach to calculate a single SED. AGNNES performs this calculation much faster than the original code: it is  $\sim 400000\times$ faster for the RIAF  and $\sim 500\times$ faster for the jet. We achieved a substantial speed-up allowing us to fit some sources (M87, NGC 4261, NGC 315). We constrained the better distribution of the parameters (see section \ref{sec:model}) to fit the SED for each object using the MCMC method. AGNNES is very accurate with respect to the fiducial method.

We built AGNNES for physical systems with fitting parameters inside its training range (see section \ref{subsec:data}). A possible future improvement is to generate more data for smaller BH masses and lower accretion rates to enhance the AGNNES training dataset. Our original code did not take into account a nonthermal leptonic population, and this can be an improvement in the future.

SED calculations are very time expensive if you need to take into account some iterative procedures as calculating Inverse Compton emission. Our work achieved a vast speed-up of some thousand times in the calculation of a single SED. This is a demonstration of the power of deep learning algorithms for astrophysical problems.

\section*{Acknowledgements}
We used \texttt{Python} \citep{python2007, python2011} to organise all SED data and to make all figures. In this work we used several packages as \texttt{pandas} \cite{pandas}, \texttt{NumPy} \citep{numpy}, \texttt{SciPy} \citep{scipy} and \texttt{Matplotlib} \citep{matplotlib}.

Figure \ref{fig:arq1} was built using the software \texttt{Nn-svg} \citep{Lenail2019}.

We acknowledge useful discussions with Jo\~ao Paulo Pe\c{c}anha Navarro, Felipe Lucas Gewers, Raniere de Menezes, Amanda Rubio and Stephane V. Werner. 
This work was supported by FAPESP (Funda\c{c}\~ao de Amparo \`a Pesquisa do Estado de S\~ao Paulo) under grants 2016/24857-6, 2017/01461-2 and 2019/10054-7. 
We gratefully acknowledge the support of NVIDIA Corporation with the donation of the Quadro P6000 GPU used for this research.

\section*{Supporting information}

The LLAGN SED fitting code will be made publicly available on github upon publication of this manuscript.


\bibliographystyle{mnras}
\bibliography{refs} 



\appendix

\section{Observational data}\label{app:obs-data}

The observational data of the modelled galaxies are available in tables \ref{tab:M87}-\ref{tab:NGC315}. Data for M87 were extracted from \cite{Prieto2016} and data for NGC315 and NGC4261 were the same as presented by \cite{DeMenezes2020}.

\input{tables.tex}

\newpage

\section{Parameter distribution}\label{app:distro}

\begin{figure*}
    \centering
    \includegraphics[width=\linewidth]{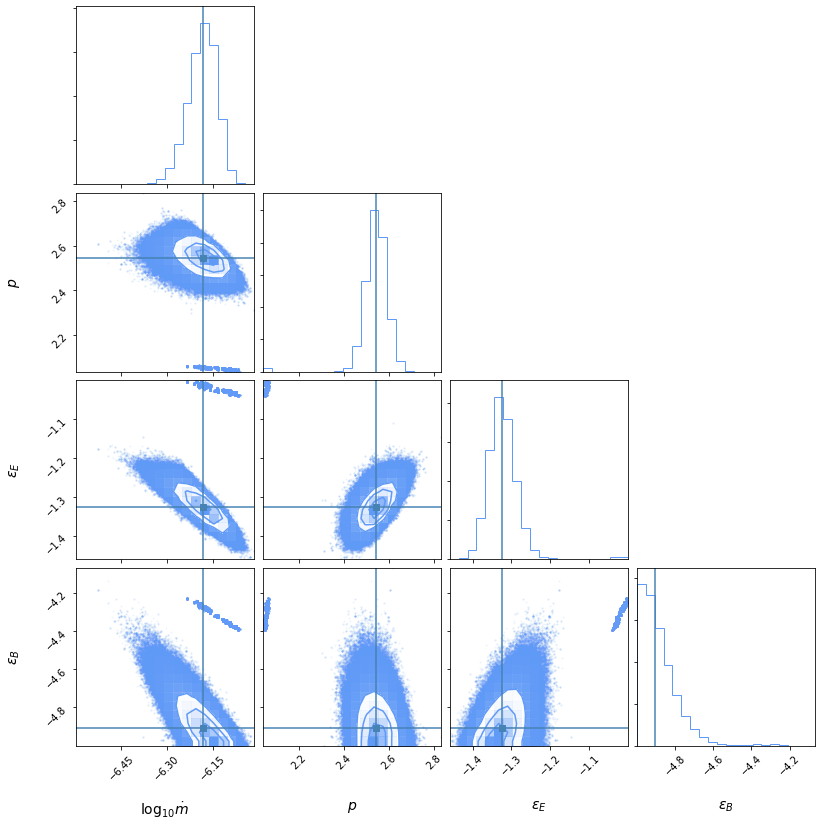}
    \caption{Corner plot for the fit parameters of M 87. In the main diagonal were plotted the \textit{posteriori} distribution of the parameters. The other plots show the correlation between each pair of variables. }
    \label{fig:87-fit}
\end{figure*}

\begin{figure*}
    \centering
    \includegraphics[width=\linewidth]{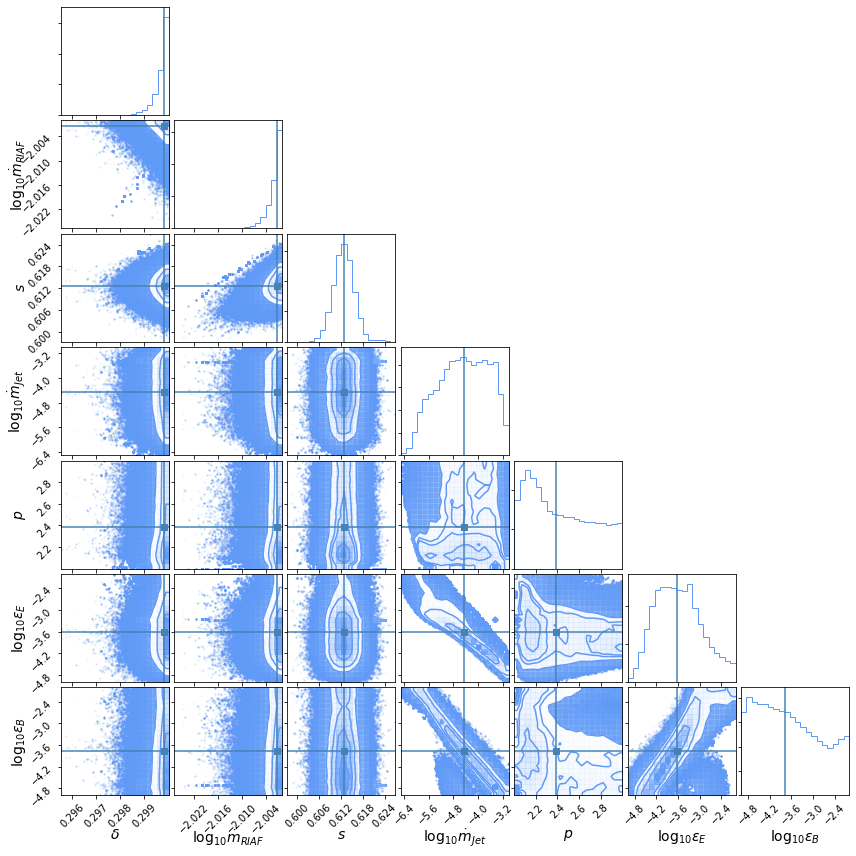}
    \caption{Corner plot for the fit parameters of NGC 4261. In the main diagonal were plotted the \textit{posteriori} distribution of the parameters. The other plots show the correlation between each pair of variables.}
    \label{fig:4261-fit}
\end{figure*}

\begin{figure*}
    \centering
    \includegraphics[width=\linewidth]{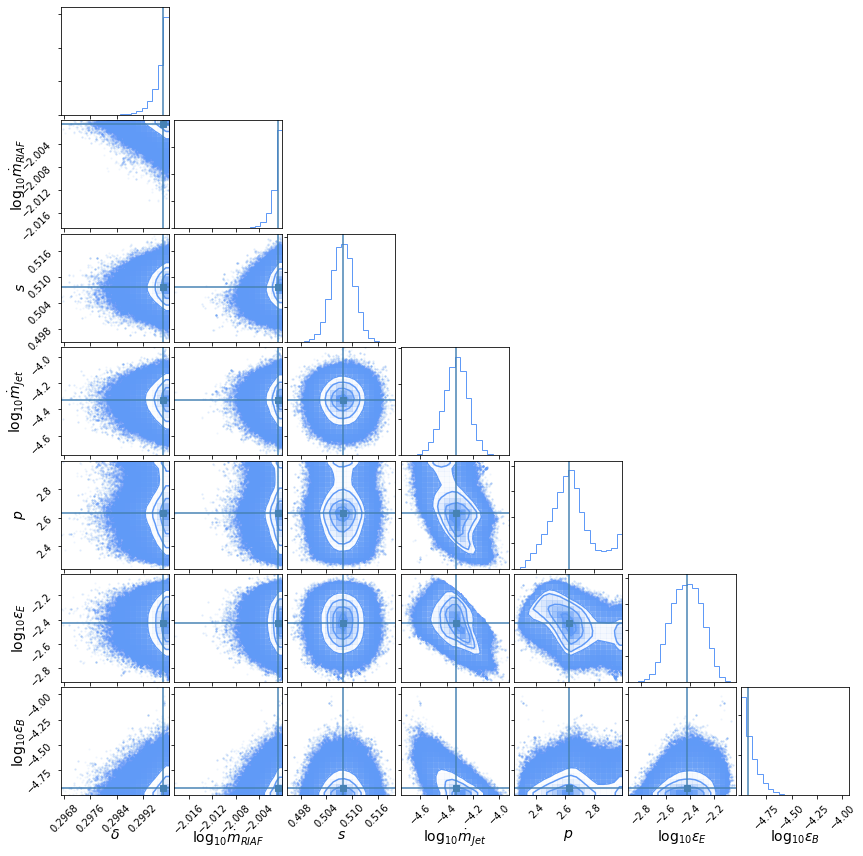}
    \caption{Corner plot for the fit parameters of NGC 315. In the main diagonal were plotted the \textit{posteriori} distribution of the parameters. The other plots show the correlation between each pair of variables.}
    \label{fig:315-fit}
\end{figure*}

In figures \ref{fig:87-fit}-\ref{fig:315-fit} we present the posterior distributions of our free parameters from the SED fits in figures \ref{fig:M87-sed}-\ref{fig:315-sed}. The plot was done using the \texttt{Python} package \texttt{corner} \citep{Foreman-Mackey2016}. The values presented in table \ref{tab:results} are the mean of the distribution and the uncertainties reported are correspondent to $1 \sigma$. We showed the parameters in the following order: $\delta$, $\dot{M}_0$, $s$, $\dot{M}_j$, $p$, $\epsilon_E$, $\epsilon_B$ -- the first three from the RIAF modeling, and the last four from the jet modeling.

\section{Neural network diagnostics}\label{app:nn}

\begin{figure*}
    \center
    \subfigure[NN architecture for RIAF]{\includegraphics[width=.8\linewidth]{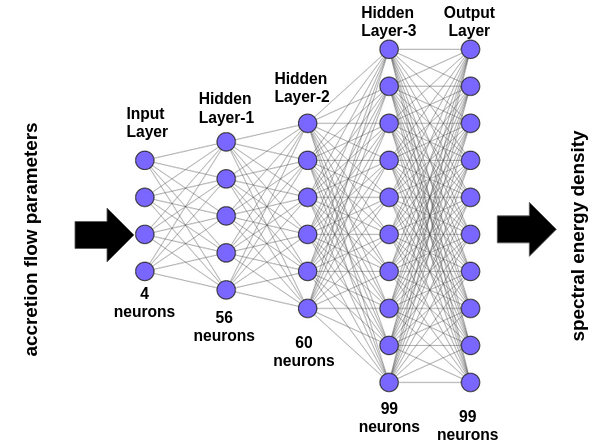}}
    \qquad
    \subfigure[NN architecture for Jet]{\includegraphics[width=.8\linewidth]{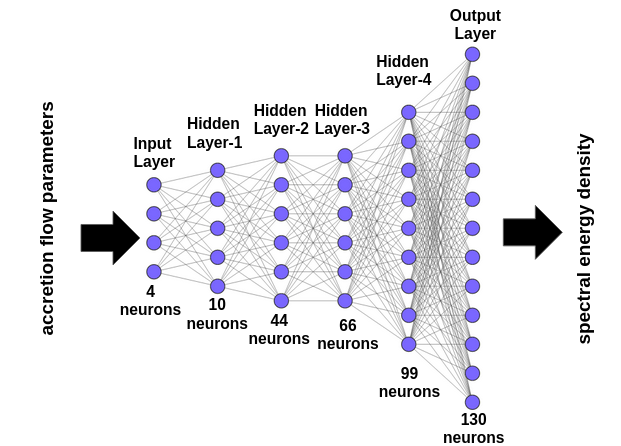}}
    \caption{The architecture (a) is the one used to predict the emission lines from RIAF mode and (b) to predict the emission line from jets.}
    \label{fig:arq1}
\end{figure*}

\begin{figure*}
    \centering
    \includegraphics[width=\linewidth]{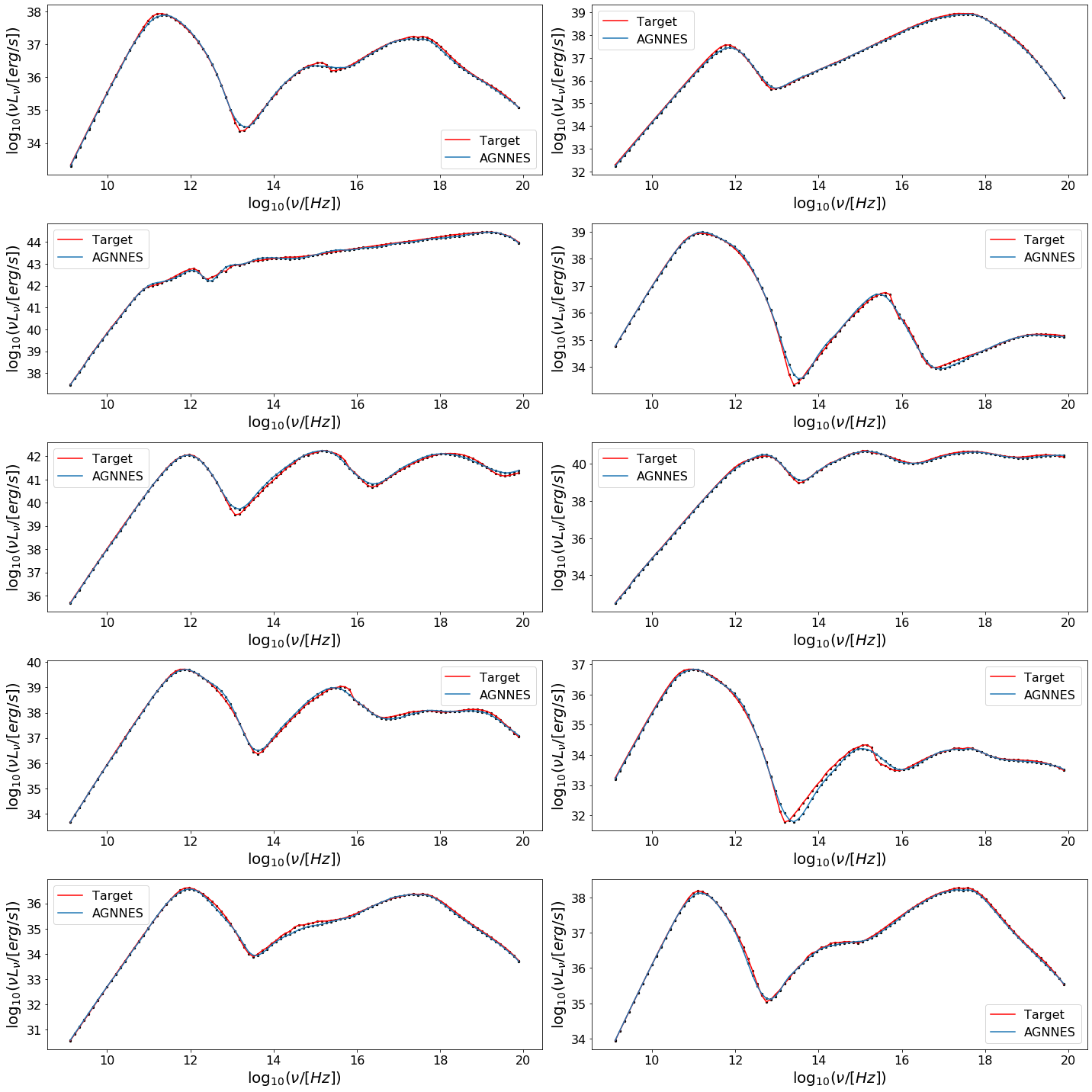}
    \caption{Plots comparing AGNNES output with the expected value for the RIAF component.}
    \label{fig:nn-RIAF}
\end{figure*}

\begin{figure*}
    \centering
    \includegraphics[width=\linewidth]{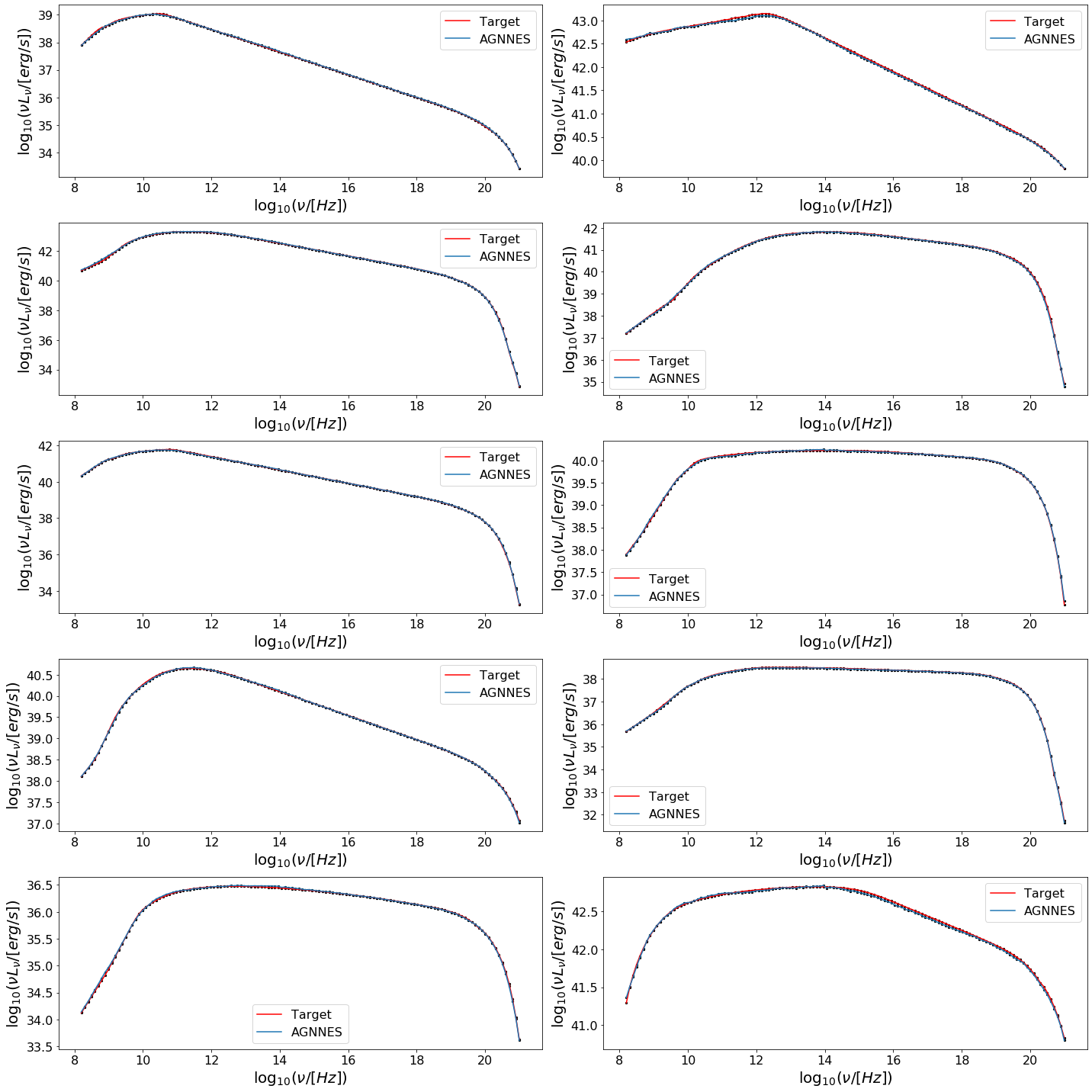}
    \caption{Plots comparing AGNNES output with the expected value for the jet component.}
    \label{fig:nn-jet}
\end{figure*}

We built AGNNES with two different NNs: one for the RIAF component, another for the jet. Figures \ref{fig:nn-RIAF}-\ref{fig:nn-jet} show how close is AGNNES prediction to the original calculation. \ref{fig:nn-RIAF} is the RIAF component and \ref{fig:nn-jet} is the jet component. In these plots, we are using some SEDs from the validation sample. 

\begin{figure*}
    \centering
    \includegraphics[width=\linewidth]{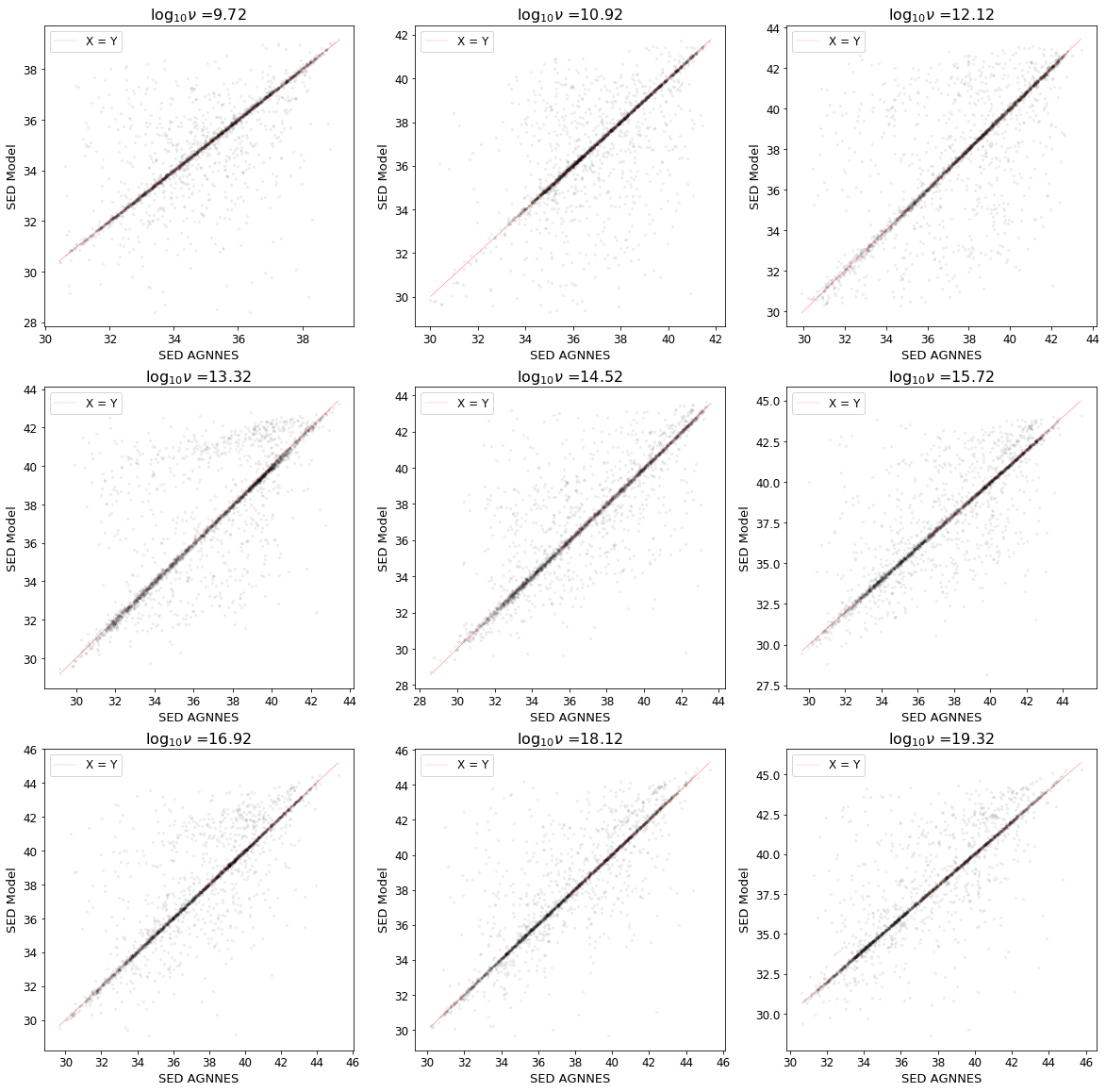}
    \caption{From the validation data set, we selected frequencies and plotted the SED value generated by the original code and the SED predicted by AGNNES for the RIAF component, respectively y-axis and x-axis. The perfect fit is the thin curve $y = x$ plotted in red.}
    \label{fig:nn-nu-RIAF}
\end{figure*}

\begin{figure*}
    \centering
    \includegraphics[width=\linewidth]{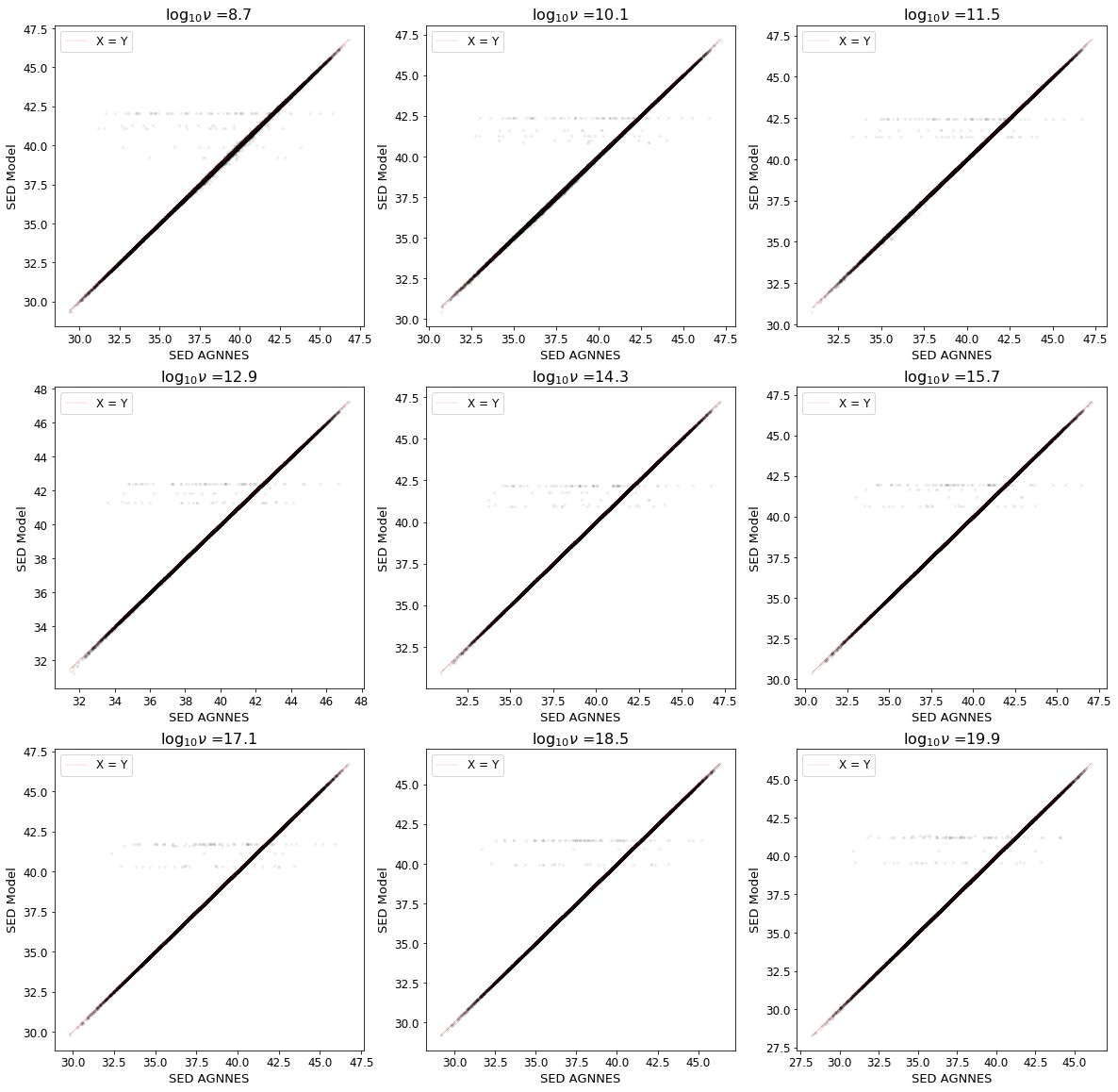}
    \caption{From the validation data set, we selected frequencies and plotted the SED value generated by the original code and the SED predicted by AGNNES for the jet component, respectively y-axis and x-axis. The perfect fit is the thin curve $y = x$ plotted in red. }
    \label{fig:nn-nu-jet}
\end{figure*}

For \ref{fig:nn-nu-RIAF} and \ref{fig:nn-nu-jet} we plotted the original value of SED against AGNNES's prediction for several frequencies, respectively the RIAF and jet components. The perfect result should be a straight line $y = x$, our results were very close to this. The further the point is from the line $x = y$, the higher is the error in the NN prediction. Some of the simulations are not physically correct, so these are underrepresented since the model learns mostly from physically accurate simulations. Another possibility to improve the results of the  $x = y$ is to feed more data to the model with every emission line shape equally represented. Finally, statistical learning is the base of NN models, so they learn based on an error function and its derivatives. Even NNs with high accuracies have an error associated. NNs with accuracy equals to $1$ are often overfitted and cannot generalize well. 

\bsp	
\label{lastpage}
\end{document}

%% file: definitions.tex
\usepackage{newtxtext,newtxmath}
\usepackage[T1]{fontenc}
\usepackage{ae,aecompl}
\usepackage{graphicx}	
\usepackage{amsmath}	
\usepackage{dblfloatfix}
\usepackage{color,soul}
\usepackage{indentfirst}
\usepackage{url}
\usepackage{scalerel}
\usepackage{placeins}
\usepackage{subfigure}
\usepackage[textwidth=1.5cm,shadow]{todonotes}	
\usepackage[normalem]{ulem}    
\usepackage{color,soul}
\usepackage{wrapfig}	

\newcommand{\orcid}[1]{\href{#1}{\includegraphics[scale=0.04]{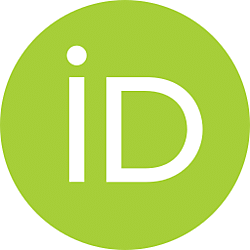}}}

%% file: tables.tex
\begin{table}
    \centering
    \begin{tabular}{lcc}
        \hline
        $\nu$ (Hz) & $\nu L_{\nu}$ (erg s$^{-1}$) & Reference \\
        \hline
        1.60E+09 & 1.87E+038 & \cite{Giovannini1990}\\
        8.40E+09 & 1.41E+039 & \cite{Morabito1986} \\
        2.20E+10 & 2.48E+039 & \cite{Junor1995}    \\
        8.60E+10 & 4.43E+039 & \cite{Lee2008}      \\
        1.00E+11 & 1.61E+040 & \cite{Lonsdale1998} \\
        2.30E+11 & 7.25E+040 & \cite{Doeleman2012} \\
        2.60E+13 & 1.09E+041 & \cite{Whysong2004}  \\
        2.80E+13 & 1.51E+041 & \cite{Perlman2001}  \\
        2.47E+14 & 1.28E+041 & \cite{Prieto2016}   \\
        3.32E+14 & 1.39E+041 & \cite{Prieto2016}   \\
        3.70E+14 & 7.55E+040 & \cite{Prieto2016}   \\
        4.99E+14 & 6.81E+040 & \cite{Prieto2016}   \\
        6.32E+14 & 8.40E+040 & \cite{Prieto2016}   \\
        8.93E+14 & 5.15E+040 & \cite{Prieto2016}   \\
        1.10E+15 & 4.53E+040 & \cite{Prieto2016}   \\
        1.27E+15 & 7.40E+040 & \cite{Prieto2016}   \\
        1.36E+15 & 4.73E+040 & \cite{Prieto2016}   \\
        2.06E+15 & 2.75E+040 & \cite{Prieto2016}   \\
        2.42E+17 & 1.95E+040 & \cite{Prieto2016}   \\
        2.42E+18 & 1.79E+040 & \cite{Prieto2016}   \\
        \hline
    \end{tabular}
    \caption{SED data for M87.}
    \label{tab:M87}
\end{table}

\begin{table}
    \centering
    \begin{tabular}{lcc}
        \hline
        $\nu$ (Hz) & $\nu L_{\nu}$ (erg s$^{-1}$) & Reference \\
        \hline
        1.63E+09 & 2.40E+38 & \cite{Jones1997} \\
        5.00E+09 & 5.88E+38 & \cite{Nagar2005} \\
        8.39E+09 & 1.24E+39 & \cite{Jones1997} \\
        1.50E+10 & 6.71E+39 & \cite{Nagar2005} \\
        1.66E+13 & ($5.39\pm0.24$)E+41 & \cite{Asmus2014} \\
        2.50E+13 & ($4.79\pm1.10$)E+41 & \cite{Asmus2014} \\
        3.64E+14 & > 6.03E+39 & \cite{Ferrarese1996} \\
        4.41E+14 & > 4.60E+39 & \cite{Ferrarese1996} \\
        5.41E+14 & > 3.96E+39 & \cite{Ferrarese1996} \\
        (4.8--24)E+17 & 1.03E+41 &  \cite{Zezas2005} \\
        \hline
    \end{tabular}
    \caption{SED data for NGC 4261.}
    \label{tab:NGC4261}
\end{table}

\begin{table}
    \centering
    \begin{tabular}{lcc}
        \hline
        $\nu$ (Hz) & $\nu L_{\nu}$ (erg s$^{-1}$) & Reference \\
        \hline
        1.40E+09  & 2.86E+39  & \cite{Capetti2005} \\
        2.50E+09  & 5.28E+39  & \cite{Lazio2001}\\
        5.00E+09  & 9.11E+39 & \cite{Nagar2005}\\
        1.50E+10 & 3.68E+40 & \cite{Nagar2005} \\
        8.62E+10 & ($2.28\pm0.13$)E+41 & \cite{Agudo2014} \\
        2.29E+11 & ($3.08\pm0.23$)E+41 & \cite{Agudo2014} \\
        3.75E+13 & ($1.47\pm0.08$)E+42 & \cite{Gu2007} \\
        5.17E+13 & ($7.72\pm0.59$)E+41 & \cite{Gu2007} \\
        6.67E+13 & ($4.63\pm0.45$)E+41 & \cite{Gu2007} \\
        8.33E+13 & ($2.95\pm0.35$)E+41 & \cite{Gu2007} \\
        3.68E+14 & 2.00E+41 & \cite{VerdoesKleijn2002} \\
        5.40E+14 & 9.48E+40 & \cite{VerdoesKleijn2002} \\ 
        (4.8--24)E+17 & 4.36E+41 & \cite{Gonzalez-Martin2006} \\   
        \hline
    \end{tabular}
    \caption{SED data for NGC 315.}
    \label{tab:NGC315}
\end{table}